\documentstyle[12pt]{article}
\begin{document}
\begin{center}{\Large {\bf Nonequilibrium phase transition in the kinetic
Ising model driven by propagating magnetic field wave}}\end{center}

\vskip 1cm

\begin{center}{Muktish Acharyya}\\
{\it Department of Physics, Presidency University,}\\
{\it 86/1 College Street, Calcutta-700073, India.}\\
{\it E-mail: muktish.acharyya@gmail.com}\end{center}

\vskip 1cm

\noindent 
The two dimensional ferromagnetic Ising model in 
the presence of a propagating
magnetic field wave (with well defined frequency and 
wavelength) is studied by
Mone Carlo simulation. 
{\it This study differs from all of the earlier studies 
done so far, where the
oscillating magnetic field was considered to be uniform in space.} 
The time average
magnetisation over a full cycle (the time period)
of the propagating magnetic field acts as
the dynamic order parameter. 
The dynamical phase transition is observed.
The temperature variation of the dynamic order
parameter, the mean square deviation 
of the dynamic order parameter, the dynamic
specific heat and the derivative of the dynamic order parameter are studied.
The mean square deviation of the dynamic order
parameter, dynamic specific heat show sharp maxima near the 
transition point.
The derivative of dynamic order parameter shows sharp minimum near the
transition point. The transition temperature is found to depend 
also on the speed of propagation of the magnetic field wave.

\vskip 1 cm



\newpage

\section {Introduction:}

The nonequilibrium response of Ising ferromagnet in the presence of 
time varying magnetic field is widely studied \cite{rev}. 
Among all these dynamical responses (e.g., hysteretic response,
dynamic phase transition, stochastic resonance etc.), the nonequilibrium
dynamical phase transition is an important phenomenon\cite{rev}
and became an interesting field of research recently.
These dynamic phase transition has several similarities with that observed
in the case of equilibrium thermodynamic phase transition. The effort
in studying the
invariance of time scale (i.e., critical slowing down) \cite{ma1},
the divergence of specific heat \cite{ma1}, divergence of critical
fluctuations in energy \cite{ma2}, divergence of length scale near
the transition point \cite{rik1}, the order of the transition \cite{rik3}
established the dynamic transition as an interesting nonequilibrium
phase transition. This dynamic transition is very closely related to
the hysteretic loss\cite{ma3} and the stochastic resonance\cite{ma5}.
Experimentally the existence of dynamic transition was found
\cite{expt} in Co film
on Cu surface (at room temperature)
by surface magneto optic Kerr effect. Recently, the evidence of dynamic phase
transition was found expreimentally
\cite{robb}, in [Co(4\AA)Pt(7\AA)]$_3$
multilayer system with strong perpendicular anisotropy by applying a time
varying (sawtooth type) out-of-plane magnetic field in the presence of small
additional constant magnetic field. In this study, the dynamic phase boundary
was drawn and found similar to that obtained from the simulation 
in kinetic Ising model with analogous condition.

The dynamic phase transition is also 
observed\cite{other1,other2,other3,other4}
in other ferromagnetic models. It is studied\cite{other5}
 in the Ginzburg-Landau
model of anisotropic XY ferromagnet and different types of chaotic
behaviour is observed.
Recently, the multiple dynamic transitions
is observed\cite{hall,ma6} in anisotropic Heisenberg model. These studies
are reviewed\cite{ma7} recently.

However, all these studies, done so far for the dynamic phase transition
are made with time varying magnetic field which was uniform over the
space. No attempt has been made to study the dynamic phase transition
with magnetic field depending on both space and time. In this article,
the dynamic phase transition is studied, by 
Monte Carlo simulation\cite{book1},
 in Ising ferromagnet in the
presence of a propagating magnetic field wave.

This article is organised as follows: the next section is devoted to
describe the model and the Monte Carlo simulation method. The simulation
results are reported in section -3 and the article ends with a summary in
section -4.

\vskip 1cm

\section{Model and Simulation:}

The Hamiltonian, of an Ising model (with ferromagnetic nearest
neighbor interaction) 
defined in two dimensions (square lattice)
in the presence of a propagating magnetic field wave, can
be represented as
\begin{equation}
H = -J\sum_{<ij>} s_i s_j -  \sum_i h(\vec r, t) s_i.
\label{hm} 
\end{equation}
Here, $s_i (=\pm 1)$ is the Ising spin variable, $J (> 0)$ is 
the ferromagnetic interaction
strength and
$h(\vec r, t)$  
is the value of the propagating magnetic field wave at any time $t$
and at position $\vec r$. Here, the propagating magnetic field 
($h(\vec r, t)$) wave is represented as

\begin{equation}
h(\vec r,t) = h_0 \cos (\omega t-Ky)
\label{field}
\end{equation}
\noindent where $h_0$ is the amplitude and $\omega (= 2\pi f)$ is 
the angular frequency of the oscillating field and $K (= 2\pi/\lambda)$ 
is the wave vector. Here, $f$ is the frequency and $\lambda$ is the 
wavelength (measured in the unit of lattice spacing)
of the propagating magnetic field wave. 
The wavelength, considered here, is smaller than 
and commensurate with the lattice size ($L$).
Here, the direction of
propagation of the magnetic field wave ($h(y,t)$)
is taken along the $y$ direction only.
{\it It may be noted here, that all earlier studies of the dynamical phase 
transitions are done with oscillating (in time) but uniform
(over the space) magnetic field.} The boundary condition is taken periodic
in all directions. This completes the description of the model.

In the simulation, the system is cooled gradually from a high temperature.
Randomly selected $50\%$ up ($s_i=+1$) spins, 
is taken as the initial configuration.
Physically, this corresponds to the high temperature 
configuration of spins. 
In the cooling process,
the last spin configuration corresponding to a particular
temperature was used as the initial configuration of 
next lower temperature.
At any finite temperature $T$,
the dynamics of this system has been studied here by Monte Carlo
simulation using Metropolis single spin-flip rate 
\cite{book1}. The transition rate 
is specified as
\begin{equation}
W(s_i \to -s_i) = {\rm Min}\left[1, \exp(-{\Delta H}/k_BT)\right]
\end{equation}
\noindent where $\Delta H$ is the change in energy due to spin 
flip ($s_i \to -s_i$)
 and $k_B$ is the Boltzmann
constant.
Any lattice site is chosen randomly and the spin 
variable ($s_i$) is updated according to the
Metropolis spin flip probability.
$L^2$ such updates constitute
the unit (Monte Carlo step per spin or MCSS) of time
here. 
The instantaneous bulk magnetisation
(per site),
 $m(t) = (1/L^2) \sum_i s_i $ has been calculated. 
The time averaged (over the complete 
cycle of the propagating magnetic field wave)
magnetisation, 
\begin{equation}
Q = {1 \over {\tau}}
\oint m(t) dt, 
\end{equation}
defines the dynamic order parameter\cite{rev}. 
The frequency is $f = 0.01$ (kept fixed throughout the study). 
So, one complete cycle of the propagating
field takes 100 MCSS (time period $\tau = {1 \over f}$ = 100 MCSS). 
A time series of magnetisation $m(t)$ has been 
generated up to $2\times10^5$ MCSS. This time series contains
$2\times10^3$ (since $\tau$ = 100 MCSS) 
number of cycles of the oscillating field. Here, first $10^3$ numbers of
such transient values are discarded to get the stable values of the
dynamical quantities.
The dynamic order parameter $Q$ has been calculated over $10^3$ values.
It is checked (for a few data) that these number of samples
($N_s$) is sufficient
to get the stable values of the dynamical quantities. 
So, the statistics 
(distribution of $Q$) is based on $N_s = 10^3$ different
values of $Q$. 
To have the confidence (with these number of samples),
the mean square deviation 
(i.e., $<(\delta Q)^2> = <Q^2>-<Q>^2$) of $Q$
is also calculated and studied as a function of temperature.
It may be noted here, that values of the 
dynamic order parameter (at lower temperatures) become both positive and
negative with equal probability. Here, only the positive values of $Q$
are shown. The statistical error
($\Delta$) in calculating $Q$ may be defined as the square root of
$<(\delta Q)^2>$. The maximum error ($\Delta_{max}$) occurs near the
transition point and this reasonably indicates the critical fluctuations.

The time average dynamic energy is defined as

\begin{equation}
E = {1 \over \tau}\oint H dt.
\end{equation}

\noindent The dynamic specific heat ($C={dE \over dT}$) is also calculated.
The temperature variations of all these (above mentioned) quantities are 
studied.

Here, the temperature $T$ is measures in the unit of $J/k_B$, the field
amplitude $h_0$ and energy $E$ are measured in the unit of $J$.

\section {Results:}

To investigate the nature of the spatio-temporal 
variations of field $h(y,t)$
and the local 'strip
magnetisation', $m(y,t)$ ($=\int {{s(x,y)} \over {L}} dx$, 
where $s(x,y)=\pm 1$
is the spin variable at position (x,y)), are
studied as a function of coordinate $y$ (along the direction of 
propagation of field wave) for different times ($t$). Fig-1, shows
such plots. From the figure, the propagating nature of the field wave
and the 'strip magnetisation' is clear. Here, it may be noted that,
for a particular instant of time, the magnetic field and the 
'strip magnetisation' differ by a phase. It is observed that, this
phase difference depends on temperature of the system, 
wavelength and the frequency of the propagating magnetic field wave.
The 
systematic study of this dependence requires lot of 
computational effort and time.

The temperature variation of the dynamic order parameter is studied. This is
shown in Fig-2. For the fixed values of the amplitude, frequency and the
wavelength of the propagating magnetic field wave, it is observed that below 
a certain temperature the dynamic ordering develops
($Q \neq 0$)
 and vanishes ($Q=0$) above it.
Keeping the values of frequency ($f$) and the wavelength ($\lambda$), 
of the propagating 
magnetic field wave, fixed, if the amplitude ($h_0$)
 of the field increases the
dynamic phase transition occurs at lower temperature. For comparison,
a similar studies are done for nonpropagating (sinusoidally oscillating
 in time but uniform over the space) magnetic field with same
frequency and amplitude. 
This clearly indicates that the dynamic transition occurs at different
higher temperatures than observed in the case of a propagating field.

These dynamic transition temperatures can be estimated by studying the
temperature variations of the mean square fluctuations
($<\delta Q^2>$) of the dynamic order parameter $Q$. These results are
shown in Fig-3. Here, the $<\delta Q^2>$ shows very sharp maximum, indicating
the dynamic transition temperature. From this one can estimate the
maximum error ($\Delta_{max}$) involved in statistical calculation for
the dynamic order parameter $Q$.
For propagating magnetic
field wave of $f=0.01$ and $\lambda = 25$, the dynamic phase transitions
(indicated by the maxima of $<\delta Q^2>$)
occur at $T=1.50$ and $T=1.88$ for the field amplitudes 
$h_0=0.5$ and $h_0=0.3$
respectively. Here also, for comparison, the similar studies are done in the
case of nonpropagating magnetic field. Here, for $f=0.01$ the dynamic 
phase transitions occur at $T=1.68$ and $T=1.94$ for $h_0=0.5$ and
$h_0=0.3$ respectively.

The derivative (${{dQ} \over {dT}}$) of the dynamic order parameter $Q$
is calculated by central difference formula\cite{book2}

\begin{equation}
{{dQ} \over {dT}} = {{Q(T+\Delta T) - Q(T-\Delta T)} \over {2\Delta T}}.
\end{equation}

\noindent In the simulation, the system was being cooled from a high
temperature (random spin configuration) to a certain temperature slowly
in the step $\Delta T = 0.02$. It may be noted here that the error in
calculating the derivative numerically by this 
central difference formula is $O({(\Delta T)^2})$\cite{book2}. 
So, the error
involved is of the order of 0.0004.
 The temperature variation of the derivative
of the dynamic order parameter is studied and the results are shown in
Fig-4. Here, the derivative shows very sharp minimum, indicating the
dynamic phase transition temperature.
For propagating magnetic
field wave of $f=0.01$ and $\lambda = 25$, the dynamic phase transitions
(indicated by very sharp minima of ${{dQ} \over {dT}}$)
are observed to occur at $T=1.50$ and $T=1.88$ 
for the field amplitudes $h_0=0.5$ and $h_0=0.3$
respectively. For a comparison, the similar studies are done in the
case of nonpropagating magnetic field. Here, for $f=0.01$ the dynamic 
phase transitions occur at $T=1.68$ and $T=1.94$ for $h_0=0.5$ and
$h_0=0.3$ respectively.

The dynamic specific heat ($C$) is calculated from the derivative 
(${{dE} \over {dT}}$) of dynamic energy ($E$). Here also, the derivative
is calculated by using central difference formula (described above).
The results are shown in Fig-5. The specific heat becomes maximum near
the dynamic transition point indicating the dynamic transition 
independently. Here, the term {\it independently} means the following: 
Here, the dynamic phase transition is studied and the transition temperature
is estimated from two types of quantities. One is dynamic order parameter
$Q$ and its derivatives (${{dQ} \over {dT}}$), moments ($<\delta Q^2>$) etc.
These depend directly on $Q$.
Another quantity is dynamic specific heat ($C={{dE} \over {dT}}$), which is 
not directly related to $Q$.
For propagating magnetic
field wave of $f=0.01$ and $\lambda = 25$, the dynamic phase transitions
(indicated by the maxima of $C={{dE} \over {dT}}$)
occur at $T=1.50$ and $T=1.88$ for the field amplitudes 
$h_0=0.5$ and $h_0=0.3$
respectively. Here also, for comparison, the similar studies are done in the
case of nonpropagating magnetic field. For $f=0.01$ the dynamic 
phase transitions occur at $T=1.68$ and $T=1.94$ for $h_0=0.5$ and
$h_0=0.3$ respectively.

The dependence of the dynamic phase transition, on the speed
of propagation of the
propagating magnetic field, is studied briefly. Here, for $f=0.01$,
$h_0=0.5$ the temperature variations of the dynamic order parameters
for $\lambda=25$ and $\lambda=50$ are studied. The results are shown
in Fig-6. It is observed that the dynamic transition occurs at higher
temperature for higher speed ($v=f\lambda$) of propagation of the 
propagating magnetic field.

The dynamical transition temperature $T_c$ is measured here for a
system of linear size $L=100$. The systematic finite size analysis
is not yet done. However, few results are checked for smaller 
(say $L=50$) system sizes. No appreciable change in $T_c$ was observed.

\section {Summary:}

The dynamical response of two dimensional Ising ferromagnet in
presence of a propagating magnetic field wave is studied by Monte 
Carlo simulation. A dynamical phase transition is observed. This
dynamical phase transition is observed from the studies of the
temperature variations of the dynamic order parameter, the derivative
of the dynamic order parameter, the mean square deviation of the
dynamic order parameter and the dynamic specific heat. 
All these studies indicate the dynamic phase
transition and the transition temperatures are estimated. 

For comparison
the dynamic transition is also studied for a nonpropagating (sinusoidally
oscillating in time but uniform over space) magnetic field. It is
observed that the dynamic transition temperatures are different from that
observed in the case of propagating magnetic field wave. It is observed,
from figures 3, 4 and 5 that the propagating field wave causes the dynamical
phase transition at lower temperature than that obtained from a 
non-propagating field of same amplitude. One may argue that since the
propagating magnetic field makes, the strip magnetisation, a wave-like
structure, the value of $Q$ will be less than that for a non-propagating
field of same amplitude and frequency at the same temperature. This would
govern the transition to take place at lower temperature.  

Here, the dependence of the transition temperature on the speed of
propagation of the propagating magnetic field wave is studied briefly
and it is observed that the transition takes place at higher temperature
for the higher value of the speed of propagation. One may try to understand
this fact in the following way: the increasing wavelength (or speed for
a fixed frequency) simply makes the field more nearly homogeneous,
approaching the infinite wavelength spatially homogeneous limit. It does
appear that the transition for propagating field (with $h_0=0.5$) 
has shifted from $T=1.50$ to that obtained
for approximately spatially homogeneous case, i.e., $T=1.68$.

The present observations, based on the Monte Carlo simulation, 
are reported here briefly.
The dynamical phase boundary for propagating magnetic field wave is
yet to be sketched and the dependence of the phase boundary on the
frequency and wavelength of the propagating wave has to be determined. 
The finite size analysis and the detailed study
of the behaviour of phase difference between propagating magnetic field 
wave and 'strip magnetisation' have to be done.
It requires lot of computational efforts and will
be reported later. The nonequilibrium dynamic phase transition
in Ising ferromagnet, in the presence of propagating magnetic field
wave, will become challenging in near future.

\vskip 1cm

\noindent {\bf Acknowledgements:} The library facilities provided
by Calcutta University is gratefully acknowledged.

\newpage

\setlength{\unitlength}{0.240900pt}
\ifx\plotpoint\undefined\newsavebox{\plotpoint}\fi
\sbox{\plotpoint}{\rule[-0.200pt]{0.400pt}{0.400pt}}%


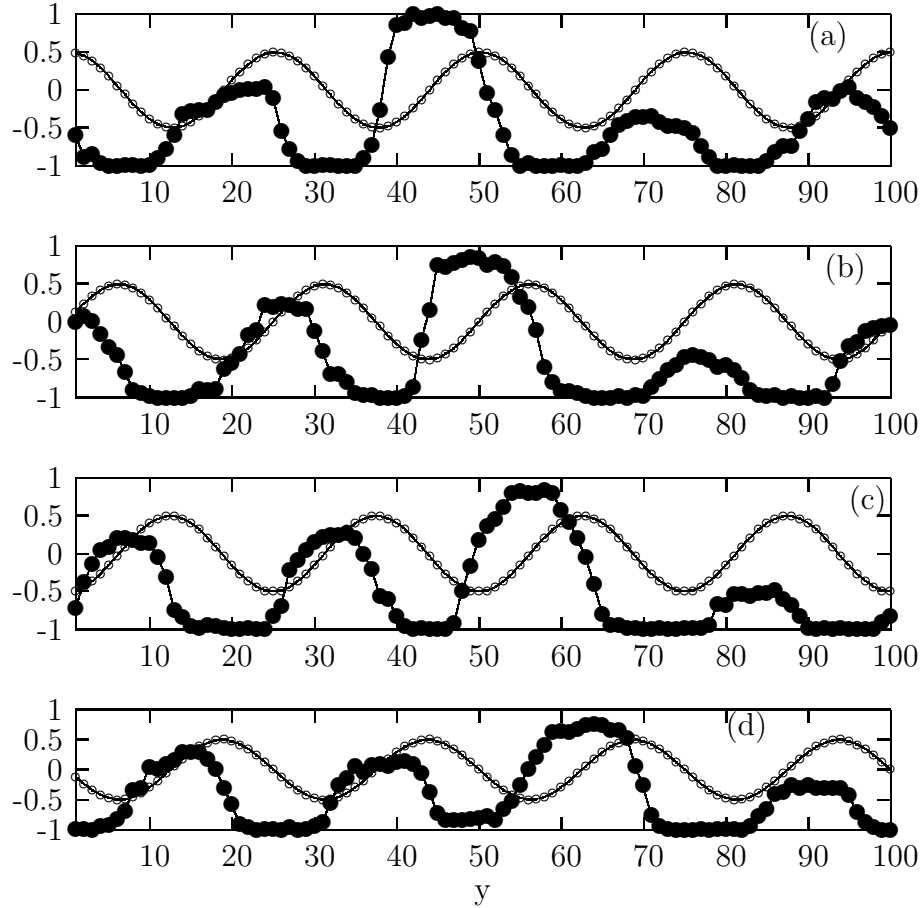
\noindent {\bf Fig-1.} The spatio-temporal variations of propagating
magnetic field wave ($h(y,t)$) and 'strip magnetisation' ($m(y,t)$)
 for $h_0=0.5$, $T=1.50$
 and $\lambda=25$.
The magnetic field and magnetisation are represented by open circles
and bullets respectively. Continuous lines joining the data points act
as guide to the eye. The plots for different times ($t$) 
are shown as follows:
(a) $t=100001$ MCSS, (b) $t=100025$ MCSS, (c) $t=100050$ MCSS and (d) $t
=100075$MCSS.

\newpage
\setlength{\unitlength}{0.240900pt}
\ifx\plotpoint\undefined\newsavebox{\plotpoint}\fi
\sbox{\plotpoint}{\rule[-0.200pt]{0.400pt}{0.400pt}}%


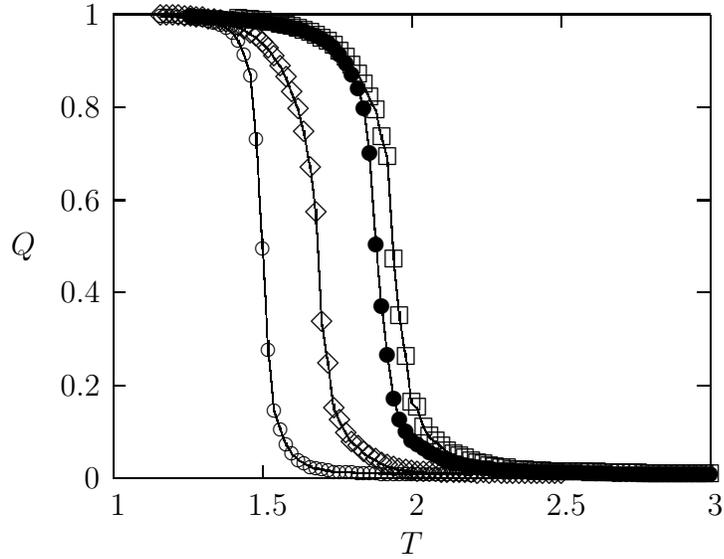
\noindent {\bf Fig.2.} The temperature ($T$) variations of dynamic order
parameter $Q$ for different types of fields. (o) for propagating
wave field with $h_0=0.5$, $\lambda=25$ (and $\Delta_{max}=0.216$), 
($\bullet$) for propagating wave field with $h_0=0.3$, 
$\lambda=25$ (and $\Delta_{max}=0.197$)
($\diamond$) for non-propagating field with
$h_0=0.5$ (and $\Delta_{max}=0.167$), 
($\Box$) for non-propagating field with $h_0=0.3$ 
and ($\Delta_{max}=0.200$). 
Here, the frequency $f=0.01$ for both type of fields.
Continuous lines just
join the data points.  

\newpage
\setlength{\unitlength}{0.240900pt}
\ifx\plotpoint\undefined\newsavebox{\plotpoint}\fi
\sbox{\plotpoint}{\rule[-0.200pt]{0.400pt}{0.400pt}}%


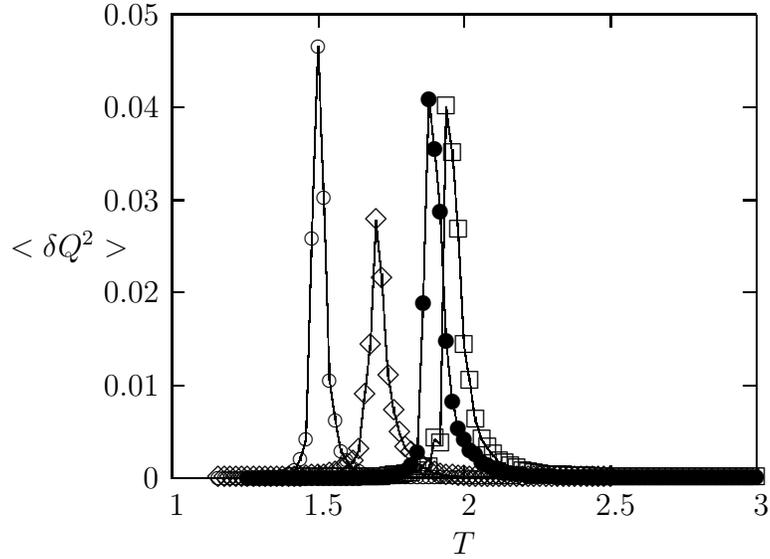
\noindent {\bf Fig.3.} The temperature ($T$) variations of mean square
deviation ($<\delta Q^2>$) of the dynamic order
parameter ($Q$) for different types of fields. (o) for propagating
wave field with $h_0=0.5$ and $\lambda=25$, 
($\bullet$) for propagating wave field with $h_0=0.3$ and $\lambda=25$, 
($\diamond$) for non-propagating field with
$h_0=0.5$, 
($\Box$) for non-propagating field with $h_0=0.3$. 
Here, the frequency $f=0.01$ for both type of fields.
Continuous lines just
join the data points.  

\newpage

\setlength{\unitlength}{0.240900pt}
\ifx\plotpoint\undefined\newsavebox{\plotpoint}\fi
\sbox{\plotpoint}{\rule[-0.200pt]{0.400pt}{0.400pt}}%


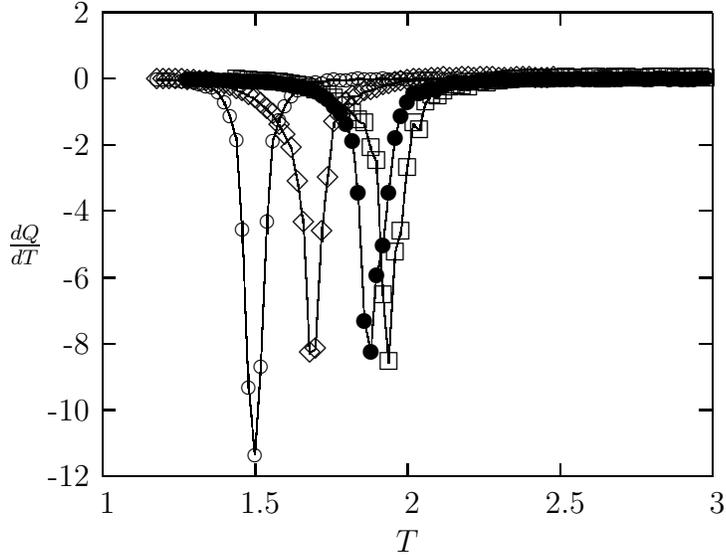
\noindent {\bf Fig.4.} The temperature ($T$) variations of the 
derivative (${{dQ} \over {dT}}$) of the dynamic order
parameter ($Q$) for different types of fields. (o) for propagating
wave field with $h_0=0.5$ and $\lambda=25$, 
($\bullet$) for propagating wave field with $h_0=0.3$ and $\lambda=25$, 
($\diamond$) for non-propagating field with
$h_0=0.5$, 
($\Box$) for non-propagating field with $h_0=0.3$. 
Here, the frequency $f=0.01$ for both type of fields.
Continuous lines just
join the data points. Here, the error involved in calculating each
data point is of the order of 0.0004.  

\newpage
\setlength{\unitlength}{0.240900pt}
\ifx\plotpoint\undefined\newsavebox{\plotpoint}\fi
\sbox{\plotpoint}{\rule[-0.200pt]{0.400pt}{0.400pt}}%


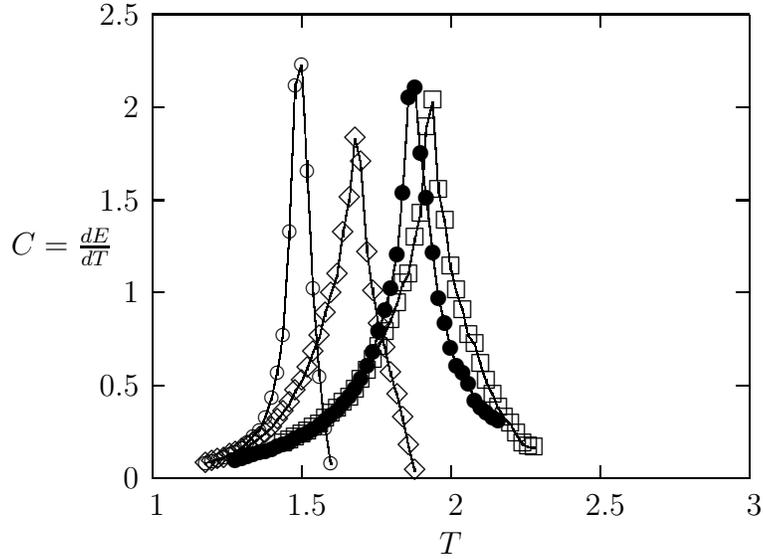
\noindent {\bf Fig.5.} The temperature ($T$) variations of the 
derivative (${{dE} \over {dT}}$) of the dynamic energy
i.e., dynamic specific heat,
 for different types of fields. (o) for propagating
wave field with $h_0=0.5$ and $\lambda=25$, 
($\bullet$) for propagating wave field with $h_0=0.3$ and $\lambda=25$,
($\diamond$) for non-propagating field with
$h_0=0.5$, 
($\Box$) for non-propagating field with $h_0=0.3$. 
Here, the frequency $f=0.01$ for both type of fields.
Continuous lines just
join the data points. Here, the error involved in calculating each 
data point is of the order of 0.0004.

\newpage
\setlength{\unitlength}{0.240900pt}
\ifx\plotpoint\undefined\newsavebox{\plotpoint}\fi
\sbox{\plotpoint}{\rule[-0.200pt]{0.400pt}{0.400pt}}%


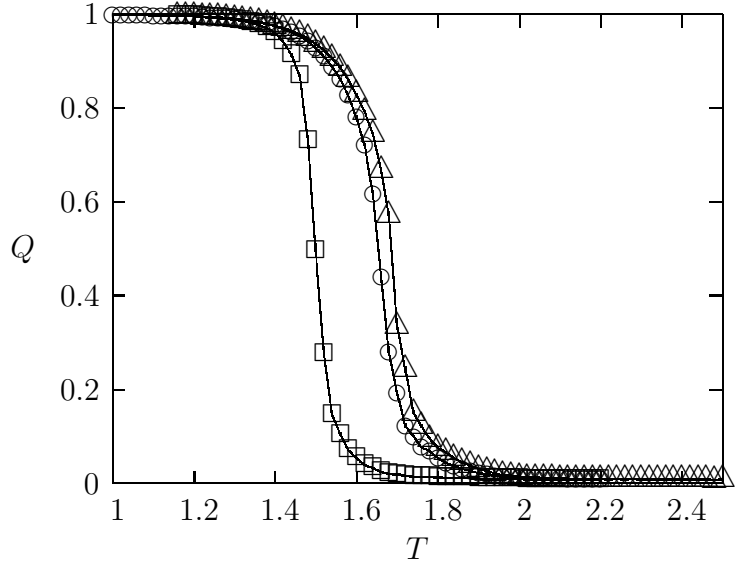
\noindent {\bf Fig.6.} The temperature ($T$) variation of dynamic order
parameter ($Q$) for propagating fields for two different 
velocities ($v=f\lambda$). 
($\Box$) 
represents $f=0.01$, $\lambda = 25$, $h_0 = 0.5$
(and $\Delta_{max}=0.216$) and (o)
represents $f=0.01$, $\lambda = 50$, $h_0 = 0.5$
(and $\Delta_{max}=0.188$). For comparison,
$Q$ versus $T$ is also plotted (represented by ($\triangle$)) for
a non-propagating oscillating magnetic field with
$f=0.01$, $h_0 = 0.5$ (and $\Delta_{max}=0.167$).
\end{document}